\documentclass[12pt]{iopart}
\usepackage{graphicx,cite}

\begin{document}

\title[]{Correlation functions and conditioned quantum dynamics in
  photodetection theory}

\author{Qing Xu, Eliska Greplova, Brian Julsgaard, and Klaus M{\o}lmer}

\address{Department of Physics and Astronomy, Aarhus University, Ny
  Munkegade 120, DK-8000 Aarhus C, Denmark}
\ead{moelmer@phys.au.dk}
\vspace{10pt}
\begin{indented}
\item[]\today
\end{indented}

\begin{abstract}
  Correlations in photodetection signals from quantum light sources
  are conventionally calculated by application of the source master
  equation and the quantum regression theorem. In this article we show
  how the conditioned dynamics, associated with the quantum theory of
  measurements, allows calculations and offers interpretations of the
  behaviour of the same quantities. Our theory is illustrated for
  photon counting and field-amplitude measurements, and we show, in
  particular, how transient correlations between field-amplitude
  measurements and later photon counting events can be accounted for
  by a recently developed theory of past quantum states of a monitored
  quantum system.
\end{abstract}

%\maketitle

\section{Introduction}
\label{sec:introduction}
The experimental observation in 1956 \cite{YEAR} by Hanbury-Brown and Twiss of temporal correlations in photodetection signals constitutes a defining moment for the field of quantum optics. While the signals measured were compatible with classical field amplitude fluctuations, in a matter of few years it became evident that temporal correlations in signals from quantum light sources, such as a single atom, would defy classical interpretation. Many scientists have since then contributed to these developments and studies of how correlations in photodetection records reveal the quantum properties of the emitter, of the light field, and even of the measurement process itself. Today, optical detection still holds a prominent role in both fundamental tests of quantum theory and in efforts to explore quantum phenomena, e.g., in precision sensing and secure communication protocols.

Correlation functions play an important role in the analysis of time dependent processes with applications ranging from demographics and finance to engineering and physics. Stochastic realizations of different signals can be characterized by, e.g., their mean values and variances while their temporal dynamics is captured by correlations between their values at different times. One may thus evaluate products of two signals at different times, $m_1(t_1) m_2(t_2)$, and define the temporal (two-time) correlation function of the signals as the average of this quantity, $\overline{m_1(t_1) m_2(t_2)}$, over different realizations of the process.
While the expression deals with classical signals $m_1(t)$ and $m_2(t)$, they may have their origin in measurements on a quantum system,
and hence we must apply quantum theory to analyze and make predictions for the signal correlations.

A central position in this research is held by Glauber's
photodetection theory \cite{Glauber, Louisell, Loudon},
which explicitly recognized that in the process of light detection,
the measurement signal is a fluctuating classical current, $i(t)$, formed by electrons that have been
excited by the absorption of photons. The mean value $\overline{i(t)}$ is proportional to the
quantum expectation value $\langle
\hat{a}^{\dagger}(t)\hat{a}(t)\rangle$, but to determine the correlation
function $\overline{i(t_1)i(t_2)}$ for $t_1 < t_2$, we must incorporate the effect of subsequent individual
photon-absorption events, leading to the normal- and time-ordered quantum expectation value
$\langle \hat{a}^{\dagger}(t_1) \hat{a}^{\dagger}(t_2)\hat{a}(t_2)
\hat{a}(t_1)\rangle$ of the field creation and annihilation operators.

Glauber's photodetection theory is formulated in the Heisenberg
\textquotedblleft operator picture\textquotedblright\ of quantum mechanics, and the evaluation of correlation functions
traditionally applies the master equation and the quantum
regression theorem for the evaluation of the observables of the light emitting system \cite{Cohen-Tannoudji, Gardiner}.
In this article we show how photodetection theory can also be naturally formulated within the general quantum theory
of measurements, which evaluates the the joint probability distribution $P(m_1,t_1; m_2, t_2)$, that photodetection signals acquire the values $m_1(t_1)=m_1$ and $m_2(t_2)=m_2$, by careful evaluation of the back action on the emitter state due to the first measurement and its subsequent evolution until the second one. When one knows the joint probability distribution, the correlation function follows by
\begin{equation}
\label{eq:Correlation_general}
\overline{m_1(t_1) m_2(t_2)} = \sum_{m_1,m_2} (m_1 m_2) P(m_1,t_1;m_2,t_2),
\end{equation}
where sums can be interchanged by integrals in the case of continuous
spectra of measurement outcomes.

The quantum measurement theory approach accounts explicitly for the back action on the state due to measurements on
the system in a manner similar to the stochastic quantum trajectory description
\cite{CARMICHAEL}. Thus, it both reproduces the results of the quantum regression theorem and sheds light on the origin of the temporal correlations in optical
detection records. We demonstrate here that the recent
past quantum state formalism \cite{PQS} provides a similar back action, but on the state \emph{prior to} a measurement event and thus offers a \textquotedblleft symmetric
intuition\textquotedblright\ to the inherently asymmetric case of cross correlations
between intensity and amplitude measurements.

The article is organized as follows:
In section~\ref{sec:Traditional_theory}, we introduce the master
equation and recall how the quantum regression theorem enables the calculation of
photodetection correlation functions. In
section~\ref{sec:general-quant-meas-theory} we review some aspects of
the general quantum theory of measurements that allows an alternative
formulation of correlation function calculations, which we extent to
the past quantum state formalism in section~\ref{sec:ampl-intens-PQS}. In
section~\ref{sec:Three-level-atoms}, we use this alternative
formulation to determine and explain interesting correlation functions
between intensity and amplitude measurements of light emitted by two-
and three-level atoms, and section~\ref{sec:conclusion} concludes the
article.

\section{Correlation functions in  photodetection theory}
\label{sec:Traditional_theory}

\subsection{The master equation and the quantum regression theorem}

%In photon counting experiments, the intensity of a light field is measured by the rate of detector clicks, each associated with the absorption of a photon from the field. When the experimental result are averaged over many independent experiments or (for stationary processes) over time, mean signals and multi-time coincidences are determined by Glauber's photodetection theory and the source master equation.

A light-emitting quantum system can be described by a density matrix
$\rho$ which, under the assumption of the Born--Markov approximation,
i.e.,~in the case of weak coupling to memoryless reservoir modes, obeys
a linear master equation,
\begin{equation}
\dot{\rho}= \mathcal{L} \rho, \label{MasterEq}
\end{equation}
where
\begin{equation} \label{lindblad} \mathcal{L} \rho =
  \frac{1}{\mathrm{i}\hbar}[H,\rho] + \sum_n \hat{C}_n \rho
  \hat{C}_n^\dagger - \frac{1}{2}(\hat{C}_n^\dagger \hat{C}_n \rho +
  \rho \hat{C}_n^\dagger \hat{C}_n).
\end{equation}
$H$ is the Hamiltonian of the system (possibly driven by
time-dependent external classical fields) while the operators
$\hat{C}_n$ account for dissipative couplings to the environment of
the system.

If the density matrix elements $\rho_{ij}$ are arranged in a vector,
$\mathcal{L}$ can be represented as a matrix, and if this matrix has
constant coefficients, equation (\ref{MasterEq}) is formally solved by
matrix exponentiation. The propagator of the master equation is a
little more complicated than a simple matrix exponential if the
equation is time dependent, but we
shall nonetheless denote it by the symbol
$\mathrm{e}^{\mathcal{L}\tau}$ and use the expression
\begin{equation}
\rho(t+\tau)=\mathrm{e}^{\mathcal{L}\tau}[\rho(t)] \label{MasterEqEvo}
\end{equation}
for the solution to equation (\ref{MasterEq}).

As stated in the Introduction, Glauber's photodetection theory
reflects how a detection signal is obtained by absorption processes in
the detector. The average product of intensities and amplitudes measured at
different times is thus formally related to the expectation value of
a normal ordered product of field creation and annihilation operators,
$\hat{a}^\dagger (t)$ and $\hat{a}(t')$. The mean intensity at time
$t$ is given by $\langle \hat{a}^\dagger (t) \hat{a}(t)\rangle$,
while the intensity--intensity correlation function, i.e., the expectation value
of the product of intensity measurements at two times $t_1 < t_2$ is
$G^{(2)}( t_1,t_2) = \langle
\hat{a}^\dagger(t_1)\hat{a}^\dagger(t_2)\hat{a}(t_2)\hat{a}(t_1)\rangle$, where the superscript $(2)$ refers to the quadratic dependence on intensity.

Like in classical electrodynamics, the electric and magnetic fields can be expressed as functions of the
charge and current distributions, and in \cite{KIMBLE} it is described in detail how the field creation and annihilation
operators develop components proportional to the raising and lowering dipole operators
of the light emitting system. For a transition in an
atom between a definite pair of excited and ground states $|e\rangle$
and $|g\rangle$, the atomic emission of photons  with rate $\gamma$ is represented by
the operator $\hat{C}=\sqrt{\gamma}|g\rangle\langle e|$ in
(\ref{lindblad}), and hence the relevant field annihilation and creation operators,
$\hat{a}$ and $\hat{a}^{\dagger}$, are replaced by the atomic transition operators
$\hat{\sigma}=|g\rangle \langle e|$ and
$\hat{\sigma}^{\dagger}=|e\rangle \langle g|$. The two-time
intensity--intensity correlation function is thus proportional to the
correlation function of atomic operators,
\begin{equation} \label{G2}
G^{(2)}( t_1,t_2)\propto \langle \hat{\sigma}^{\dagger}(t_1)  \hat{\sigma}^{\dagger}(t_2) \hat{\sigma}(t_2)\hat{\sigma}(t_1)\rangle.
\end{equation}
The two-time expectation values in (\ref{G2}) is on the
form $\langle \hat{O}_1(t)
\hat{O}_2(t+\tau)\hat{O}_3(t)\rangle=\Tr(\hat{O}_1(t)
\hat{O}_2(t+\tau)\hat{O}_3(t)\rho(t))=\Tr(\hat{O}_2(t+\tau)\hat{O}_3(t)\rho(t)
\hat{O}_1(t))$, where the central operator $\hat{O}_2(t+\tau)$ $(=\hat{\sigma}^{\dagger}(t_2) \hat{\sigma}(t_2))$ can be expanded on the
complete set of dyadic operators $(|j\rangle \langle i|)(t+\tau)$. The expectation values of these operators
are the density matrix elements, $\rho_{ij} = \langle
(|j\rangle \langle i|)\rangle$, which obey the linear set of
coupled equations defined by the master equation (\ref{MasterEq}). The same is, however, not
true for the operators $(|j\rangle\langle i|)$ themselves, and noise
operator terms must be added to the equations to ensure the
preservation of commutation and uncertainty relations
\cite{Cohen-Tannoudji, Gardiner}.
In the Markov approximation, however, these noise operators are uncorrelated with all system
observables at earlier times, and this implies that two- and
multi-time operator correlation functions of the form, $\langle \hat{O}_1(t)
(|j\rangle\langle i|)(t+\tau) \hat{O}_3(t)\rangle$, indeed, evolve with the time argument $\tau$
according to the same linear set of coupled equations as $\langle
(|j\rangle \langle i|)(t+\tau)\rangle$, \textit{i.e}., as $\rho_{ij}(t+\tau)$,
as long as $\tau > 0$ and $\hat{O}_1(t)$ and $\hat{O}_3(t)$ represent operators, evaluated at earlier
times. This is the Quantum Regression Theorem \cite{Gardiner, CARMICHAEL1}, and it offers an effective means to
determine correlation functions of operators for
Markovian open quantum system.

For comparison with experiments, we are interested in
steady state dynamics, and hence only in the dependence of the
correlation functions on the time difference $\tau$. We thus assume
that $\rho(t)= \rho^{\mathrm{st}}$, the steady state solution of the
master equation, and, for $\tau > 0$, the quantum regression theorem yields the formal
solution \cite{Gardiner, CARMICHAEL1}
\begin{equation}
\langle\hat{O}_{1}(t)\hat{O}_{2}(t+\tau)\hat{O}_{3}(t)\rangle=\mathrm{Tr}%
\left(  \hat{O}_{2}\mathrm{e}^{\mathcal{L}\tau}[\hat{O}_{3}\rho^{\mathrm{st}}\hat{O}%
_{1}]\right). \label{QuantumReg}
\end{equation}
The master equation is linear and preserves the trace and hence, for any
matrix $\mu$,
$\mathrm{e}^{\mathcal{L}\tau}[\mu] \rightarrow
\mathrm{Tr}(\mu)\rho^{\mathrm{st}}$ for $\tau \rightarrow \infty$. This implies that the correlation
function (\ref{QuantumReg}) regresses from the steady state mean value
of the operator product $\langle \hat{O}_1
\hat{O}_2\hat{O}_3\rangle_{\mathrm{st}}$ at $\tau=0$ to the product of
steady state mean values $\langle \hat{O}_2\rangle_{\mathrm{st}}
\langle \hat{O}_1\hat{O}_3\rangle_{\mathrm{st}}$. In quantum optics it is convenient to normalize the
correlation functions by this product of the steady state mean values. The resulting reduced correlation functions, denoted by a lower case letter $g$, then approach unity for large $\tau$.

\subsection{Intensity--intensity and intensity--amplitude correlations}
With the choice of operators
$\hat{O}_1=\hat{\sigma}^\dagger,\hat{O}_3=\hat{\sigma}$, and
$\hat{O}_2=\hat{\sigma}^\dagger \hat{\sigma}$, the
intensity--intensity correlation function $G^{(2)}(\tau)$ (\ref{G2})
is given by (\ref{QuantumReg}). With these operators, $\hat{O}_3
\rho^{\mathrm{st}} \hat{O}_1= \rho_{ee}^{\mathrm{st}}|g\rangle\langle
g|$, i.e., the atomic ground state density operator, multiplied by the
steady state excitation probability $\rho_{ee}^{\mathrm{st}}$. In this
case, the evolution by $\exp({\mathcal{L}\tau})$ merely solves the
transient evolution of the atomic density matrix, starting from the
ground state, and since $\hat{O}_2=|e\rangle\langle e|$ is the
projection operator on the excited state, the reduced correlation function evaluates to $g^{(2)}(\tau) =
\rho_{ee}^{\mathrm{st}} \rho_{ee}(\tau)|_g/(\rho_{ee}^{\mathrm{st}})^2=\rho_{ee}(\tau)|_g/\rho_{ee}^{\mathrm{st}}$, where
$\rho_{ee}(\tau)|_g$ denotes the excited state population of a system,
evolved from the ground state at $\tau=0$. This result is readily
understood from the quantum trajectory picture \cite{CARMICHAEL,
  DALIBARD, WISEMAN}: The photon detection at time $t$ is accompanied
by an atomic quantum jump into the ground state, and the probability
to register another photon from the same atom at the later time
$t+\tau$ is proportional to the excited state population
conditioned on the first detection.
\begin{figure}[tbp]
\centering%
\includegraphics[bb=154 462 441
688,width=0.5\columnwidth,keepaspectratio]{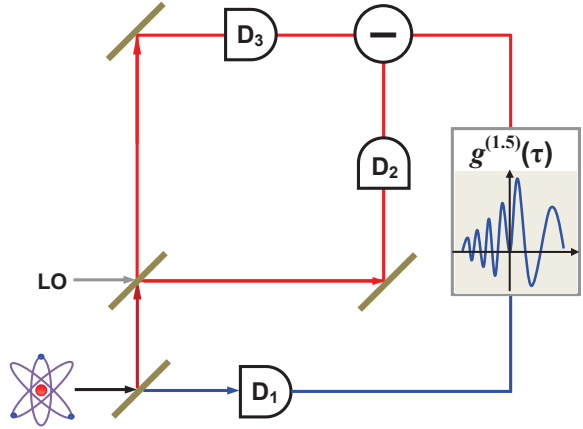}
\caption{A setup for joint photon counting and homodyne
  detection. Light transmitted by the beam splitter along the lower
  path is detected as discrete photon counting events, while the light
  reflected along the upper path is mixed with a strong local
  oscillator and the difference between the two counting signals
  represents the continuous homodyne monitoring of the emitted field
  amplitude. Averaging the noisy, time dependent homodyne signal in
  time windows around all counting events in the detector
  $\mathrm{D}_1$, we obtain the amplitude--intensity correlation
  function.}
\label{homodynedetection}
\end{figure}

The particle and the wave description of optical phenomena competed as the dominant descriptions of light until Planck and Einstein introduced the dual quantum nature of light more than a century ago.
In beautiful, more recent demonstrations of the joint wave and particle properties
of light \cite{CarmichaelREF2,CarmichaelREF4}, the radiation emitted by an atom was split in
two components, see figure \ref{homodynedetection}, for which the
photon number and the field amplitude were detected,
respectively. These experiments revealed strong temporal correlations
between the measurement outcomes. For a positive time delay, $\tau
>0$, between the counting events and the amplitude measurement, such
amplitude--intensity correlations are captured by the time- and
normal-ordered coherence function \cite{CarmichaelREF2}
\begin{equation}
g^{(1.5)}(\tau >0)=\frac{\langle\hat{\sigma}^{\dagger
}(t)(\mathrm{e}^{\mathrm{i}\varphi}\hat{\sigma}^{\dagger}(t+\tau)+\mathrm{e}^{-\mathrm{i}%
\varphi}\hat{\sigma}(t+\tau))\hat{\sigma}(t)\rangle}{\rho_{ee}^{\mathrm{st}}
(\mathrm{e}^{\mathrm{i}\varphi}\rho_{ge}^{\mathrm{st}}+\mathrm{e}^{-\mathrm{i}\varphi}\rho_{eg}^{\mathrm{st}})} ,\label{G15pos}%
\end{equation}
where $\varphi$ is the phase component measured by the homodyne
detector, and where the superscript $(1.5)$ refers to the
\textquotedblleft intensity$^{1.5}$\textquotedblright -dependence of
the function. The numerator in this expression is represented by
(\ref{QuantumReg}), with $\hat{O}_1=\hat{\sigma}^\dagger$,
$\hat{O}_3=\hat{\sigma}$, and $\hat{O}_2 =
\mathrm{e}^{\mathrm{i}\varphi}{\hat{\sigma}}^{\dagger} +
\mathrm{e}^{-\mathrm{i}\varphi}{\hat{\sigma}}$, and its behaviour can be understood from the quantum jump of the steady
state density matrix into the ground state at time $t$, and the
evaluation of the expectation value of the amplitude operator
$\hat{O}_2$ during the subsequent transient evolution of the
conditional state.

\begin{figure}[tbp]
\centering%
\includegraphics[bb=147 426 439
697,width=0.5\columnwidth,keepaspectratio]{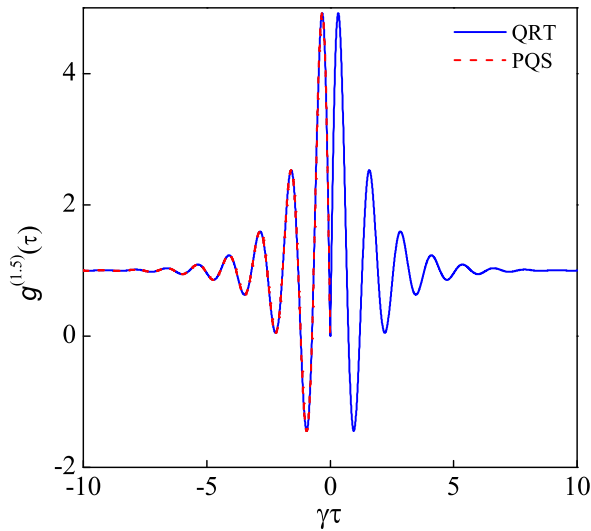}
\caption{Amplitude--intensity correlation function
  (\ref{G15pos}) and (\ref{G15neg}) for a resonantly driven two-level atom determined by the
  quantum regression theorem (blue solid line) and by the past quantum state (red dashed line) (\ref{pqs}). The
  parameters are: $\Omega =5\gamma$ and $\varphi =\pi /2$.}
\label{2levelfig}
\end{figure}

In the experiments \cite{CarmichaelREF2,CarmichaelREF4}, the homodyne amplitude signal shows correlations also with later
photon counting events. The measured correlations involve the same
operators $\hat{O}_i$ as before, but for $\tau<0$, the time- and
normal-ordering procedure leads instead to the expression \cite{CarmichaelREF3}
\begin{equation}
g^{(1.5)}(\tau <0)=\frac{\mathrm{e}^{\mathrm{i}\varphi }\langle \hat{\sigma}%
^{\dagger }(t)\hat{\sigma}^{\dagger }(t-\tau )\hat{\sigma}(t-\tau )\rangle
\newline
+\mathrm{e}^{-\mathrm{i}\varphi }\langle \hat{\sigma}^{\dagger }(t-\tau )%
\hat{\sigma}(t-\tau )\hat{\sigma}(t)\rangle }{\rho _{ee}^{\mathrm{st}}(%
\mathrm{e}^{\mathrm{i}\varphi }\rho _{ge}^{\mathrm{st}}+\mathrm{e}^{-\mathrm{%
i}\varphi }\rho _{eg}^{\mathrm{st}})}.  \label{G15neg}
\end{equation}
In the evaluation of (\ref{G15neg}), the time evolution
operator in (\ref{QuantumReg}) is  applied to products of matrices
$\rho ^{\mathrm{st}}\hat{\sigma}^{\dagger}$ and $\hat{\sigma}{\rho
  ^{\mathrm{st}}}$ that do not permit separate interpretation as
quantum states, and unlike the case of $\tau>0$, we do not have a
simple interpretation of the expression in (\ref{G15neg}) in terms of
the transient conditioned dynamics of the system. Wiseman has shown
\cite{WISEMAN2002}, however, that the weak value formalism \cite{Aharonov} for the
post selected average of a weakly perturbing measurement, indeed, accounts for
field-amplitude correlations with later count events. That formalism
is related to the recently developed theory of past quantum states \cite{PQS}, and as described below, the general quantum theory
of measurements and the past quantum state, allow both calculation
(the red dashed curve in figure \ref{2levelfig}) and interpretation of
how field-amplitude measurements are correlated with later detector
click event.

In figure \ref{2levelfig}, we show the value of the
normalized correlation function $g^{(1.5)}(\tau)$ for a two-level
atom for both positive and negative $\tau$, and it is interesting that despite the very
different formal expressions, the evaluation of (\ref{G15pos}) and
(\ref{G15neg}) yields a symmetric correlation function around
$\tau=0$. Since intensity--intensity correlation functions concern fully equivalent detection events, they have to be symmetric,
but as we shall see below, intensity--amplitude correlation functions may, indeed, be very asymmetric.

\section{Quantum measurement theory and field correlations}
\label{sec:general-quant-meas-theory}
Equation (\ref{MasterEq}) describes the average behavior of an
unobserved light emitting system, while observation of the emitted
radiation will yield a conditioned dynamics, described by the general
theory of measurements. Formally, any observation of a quantum system
is described by a positive operator valued measure (POVM), i.e., a set
of operators $\hat{\Omega}_m$, with $\sum_m \hat{\Omega}_m^\dagger
\hat{\Omega}_m = I$, where $I$ is the identity matrix \cite{ WISEMAN, Nielsen}. The operators $\hat{\Omega}_m$ account for the joint
evolution of the system and an appropriate meter system and read-out
of the meter in the states that we associate with the outcome results
$m$.  The probability to get a definite outcome $m$ is given by the
expression,
\begin{equation} \label{POVMP}
P(m)=\textrm{Tr}(\rho \hat{\Omega}_m^\dagger \hat{\Omega}_m),
\end{equation}
and the (normalized) state of the system, conditioned on this
measurement outcome, is
\begin{equation} \label{POVMS}
\rho_m= \hat{\Omega}_m \rho \hat{\Omega}_m^\dagger/P(m).
\end{equation}
If $\{\hat{\Omega}_m\}$ constitutes a complete set of orthogonal
projection operators, the POVM formalism yields the conventional Born
rule and von Neumann projection postulate.

%The average behaviour of the system, subject to the different possible measurement outcomes during a small time step $dt$, is given by
%\begin{equation} \label{POVM}
%\rho(t+dt) =   \sum_m P(m) \frac{\hat{\Omega}_m  \rho(t) \hat{\Omega}_m^\dagger}{P(m)} =  \sum_m \hat{\Omega}_m  \rho(t) \hat{\Omega}_m^\dagger.
%\end{equation}

The POVM formalism also gives access to the probability that a
measurement, described by POVM operators $\hat{\Omega}_m$ yields $m_1$
at time $t_1$ \textit{and a subsequent measurement} at $t_2$
\textit{on the same system}, described by the same or a different set
of POVM operators, yields the outcome $m_2$. Such joint probabilities
follow Bayes' rule, and can be written as
\begin{equation} \label{Bayes}
P(m_1, t_1;m_2, t_2) = P(m_1,t_1)\cdot P(m_2,t_2|m_1,t_1).
\end{equation}
Assuming steady state conditions, the unconditional probability $P(m_1)=$Tr$(\hat{\Omega}_{m_1}
\rho^{\mathrm{st}} \hat{\Omega}_{m_1}^\dagger)$ for
the first event to occur does not depend on the time argument $t_1$, and the state of the
system, conditioned on this outcome is $\rho_{m_1}(t_1) =
\hat{\Omega}_{m_1} \rho^{\mathrm{st}}
\hat{\Omega}^\dagger_{m_1}/P(m_1)$. Propagating the density matrix further, until $t_2$, by
the usual master equation with $\rho_{m_1}(t_1)$ as the initial
condition, we find
\begin{equation}
\rho_{m_1}(t_2) = \mathrm{e}^{\mathcal{L}(t_2-t_1)} \rho_{m_1}(t_1).
\end{equation}
For the joint outcome probability distribution, we thus obtain
\begin{equation} \label{jointdistribution} P(m_1,t_1;m_2,t_2) =
  \textrm{Tr}(\hat{\Omega}_{m_2}
  \mathrm{e}^{\mathcal{L}(t_2-t_1)}[\hat{\Omega}_{m_1}
  \rho^{\mathrm{st}} \hat{\Omega}^{\dagger}_{m_1}]
  \hat{\Omega}^\dagger _{m_2}),
\end{equation}
where $\mathrm{e}^{\mathcal{L}(t_2-t_1)}$ acts on the matrix
object contained within the square brackets.

Equation (\ref{jointdistribution}) provides the outcome joint
probabilities and offers a microscopic, dynamic view of a conditionally evolved quantum state.
This differs from (\ref{QuantumReg}) which directly provides the average
correlation function of the physical observables, e.g., of the field
and intensity operators. They are connected, however, by using
equation (\ref{jointdistribution}) for the calculation of weighted
mean values of the outcomes: If the labels $m_1$ and $m_2$ denote the
actual values of the physical observables measured at times $t_1$ and
$t_2$, the operator correlation function of interest acquires the form (1),

\begin{equation}
\overline{m_1m_2} = \sum_{m_1,m_2} (m_1m_2)P(m_1,t_1;m_2,t_2).
\end{equation}

\subsection{Photon counting and homodyne detection}

Restricting ourselves to the case of detection of light emitted from a
quantum system, photon counting associates two POVM operators to the
click and no-click detector events during an infinitesimal time interval $\mathrm{d}t$. The
observation of a photon from a two-level atom is described by
$\hat{\Omega}_1 = \sqrt{\gamma \mathrm{d}t}\hat{\sigma}$, while the
absence of a detector click is described, to first order in
$\mathrm{d}t$, by $\hat{\Omega}_0 = (I-\frac{\gamma \mathrm{d}t}{2}
\hat{\sigma}^\dagger \hat{\sigma})$, reproducing the respective
probabilities $\rho_{ee}\gamma \mathrm{d}t$ and $1-\rho_{ee}\gamma
\mathrm{d}t$ as well as the final atomic states after the detection
events. The atomic states are notably not orthogonal for the two field
detection outcomes. The average behavior of the system, $\rho
\rightarrow \sum_m \hat{\Omega}_m \rho \hat{\Omega}_m^\dagger$,
subject to the two outcome possibilities is exactly the one described
by the damping terms in the master equation (\ref{MasterEq}) with a single
$\hat{C}=\sqrt{\gamma}\hat{\sigma}$.

In homodyne detection, the light signal is mixed with a strong
classical field of phase $\varphi$, and the intensity of the
interfering fields is recorded by two photon counters, as shown in
figure \ref{homodynedetection}. The difference between their counting
signals is represented by a continuous quantity $x$, which in the
absence of the emitter is represented by a Gaussian distribution, $P_0(x) = \frac{1}{\sqrt{2\pi/\mathrm{d}t}}\exp(-x^2 \mathrm{d}t/2)$,
with a suitable normalization of the outcome argument $x$ \cite{WISEMAN, WISEMAN2002}. In the presence of the two-state emitter, the outcome is
governed by the family of POVM operators,
\begin{equation}
\hat{\Omega}_x = \left (\frac{\mathrm{d}t}{2\pi}\right )^{\frac{1}{4}} \textrm{exp}\left (-\frac{x^2}{4} \mathrm{d}t\right )
(1+x \mathrm{d}t\sqrt{\gamma} \mathrm{e}^{-\mathrm{i}\varphi}\hat{\sigma} -\frac{\gamma \mathrm{d}t}{2} \hat{\sigma}^\dagger\hat{\sigma}).
\end{equation}
To first order in $\mathrm{d}t$, one readily verifies the POVM
property $\int \hat{\Omega}_x^\dagger \hat{\Omega}_x \mathrm{d}x = I$
and that the average change of the state, $\rho \rightarrow \int
\hat{\Omega}_x \rho \hat{\Omega}_x^\dagger \mathrm{d}x$, reproduces the effect of the damping term
$\hat{C}=\sqrt{\gamma}\hat{\sigma}$ in (\ref{lindblad}).

%
%We have thus established that the average evolution of the conditional dynamics of the system subject to detection is given by the master equation. Conversely, it is a general theorem CITE  that operators $K_{\alpha}$ exist that obey $\sum_\alpha K_{\alpha}^\dagger K_{\alpha} = I$, such that the average density matrix evolution from $t_1$ to $t$, can be written on the form
%\begin{equation} \label{kraus}
%\rho(t) = \sum_{\alpha} K_{\alpha} \rho(t_1) K_{\alpha}^\dagger.
%\end{equation}
%We may think of the index $\alpha$ as the sequence of  $m=0,\ 1$ values corresponding to photon counting or $x$ values corresponding to homodyne detection in every interval $\mathrm{d}t$ from time $t_1$ to $t$, and the sum in (\ref{kraus}) as a sum over the corresponding quantum trajectories. For a given master equation, there exist efficient methods to identify candidate operators $K_\alpha$ that yields (\ref{kraus}) without reference to measurements CITE. We shall, however, not need explicit expressions for $K_\alpha$ in the following, but we \textit{shall} make use of the fact that the non-observed time evolution can be written as in (\ref{kraus}).

The mean intensity  measured as the average number of detection events
per time, $\overline{n}/dt$, has an expectation value given by the weighted mean of the
possible outcomes of the photon counting measurement, $\bar{n} =
\sum_{n=0,1} n\cdot P(n) = \sum_{n=0,1} n \cdot \mathrm{Tr}
(\hat{\Omega}_n \rho {\hat{\Omega}^{\dagger}}_{n}) =
\rho^{\mathrm{st}}_{ee} \gamma \mathrm{d}t$, where the last expression
applies for the two-level atom in steady state prior to the
measurement. Similarly, in homodyne detection, the one-time mean
amplitude of the field is proportional to $\langle x\rangle = \int
x\cdot P(x) \mathrm{d}x =\int x\cdot \Tr(\hat{\Omega}_x \rho
\hat{\Omega}^{\dagger}_x)\mathrm{d}x =
\sqrt{\gamma} \Tr(\mathrm{e}^{-\mathrm{i}\varphi}\hat{\sigma}\rho +
\mathrm{e}^{\mathrm{i}\varphi}\rho\hat{\sigma}^\dagger) =
\sqrt{\gamma}(\mathrm{e}^{-\mathrm{i}\varphi}\rho^{\mathrm{st}}_{eg}+\mathrm{e}^{\mathrm{i}\varphi}\rho^{\mathrm{st}}_{ge})$.

The two-time intensity--intensity correlation function is proportional
to $\sum_{n_1,n_2=0,1} n_1 n_2 P(n_1,t_1;n_2,t_2) =
P(n_1=1,t_1;n_2=1,t_2)$ with both of the POVM operators in equation
(\ref{jointdistribution}) pertaining to photon counting. Applying the
expressions for a two-level emitter, we thus obtain
$P(n_1=1,t_1;n_2=1,t_2=t_1+\tau) = (\gamma \mathrm{d}t)^2
\rho^{\mathrm{st}}_{ee} \rho_{ee}(\tau)|_{g}$, which is proportional
to $G^{(2)}(\tau>0)$, identified in the previous section. In exactly
the same way, we can determine the probability for the joint detection
of a photon at time $t_1$ and the subsequent amplitude outcome $x$ at
time $t_2$, $P(n,t_1;x,t_2)$, and we can determine the average $\sum_n
\int nx P(n,t_1;x,t_2)\mathrm{d}x$. For positive time lag
$\tau=t_2-t_1$ this yields $
\rho^{\mathrm{st}}_{ee}(\mathrm{e}^{-\mathrm{i}\varphi}
\rho_{eg}(\tau)|_g+\mathrm{e}^{\mathrm{i}\varphi}\rho_{ge}(\tau)|_g)\gamma \sqrt{\gamma} \mathrm{d}t $
which is indeed proportional to $G^{(1.5)}(\tau)$ (\ref{G15pos}).

\section{Amplitude--intensity correlations and past quantum state}
\label{sec:ampl-intens-PQS}
We shall now address the joint probability of a homodyne detector
outcome $x$ at the earlier time $t_1$ followed by the counting of a
photon at the later time $t_2$, i.e., we take
$\hat{\Omega}_{m_1}=\hat{\Omega}_x$ and
$\hat{\Omega}_{m_2}=\hat{\Omega}_n$ in (\ref{jointdistribution}),
\begin{eqnarray}
P(x,t_{1};n,t_{2}) &=&P(x,t_{1})\cdot P(n,t_{2}|x,t_{1})  \nonumber \\
&=&\Tr(\hat{\Omega}_{n}\mathrm{e}^{\mathcal{L}(t_{2}-t_{1})}[\hat{\Omega}%
_{x}\rho ^{\mathrm{st}}\hat{\Omega}_{x}^{\dagger }]\hat{\Omega}_{n}^{\dagger
}).  \label{Pxt1nt2}
\end{eqnarray}
To evaluate the amplitude--intensity correlation function, we multiply
the probability by $n$ and $x$, and we sum/integrate over their
possible values. This will retain only the $n=1$ component and the
contributions that are first order in $x$ in the expressions for
$\hat{\Omega}^{(\dagger)}_x$. Thus, even though the $\hat{\Omega}_{x}$ operators differ only infinitesimally from the identity and the back action of the $x$ measurement is weak, when we calculate the integral over $x$, only the terms $\propto\ dt\sqrt{\gamma}e^{-i\varphi}\hat{\sigma}$ and $dt\sqrt{\gamma}e^{i\varphi}\hat{\sigma}^\dagger$, \textit{cf}. (15), contribute to the correlation function average over the very nosy homodyne signal.

Rather than merely reproducing existing results with a different
method, the purpose of this work is to analyze the problem and gain
new insight into the results. We thus rewrite equation (\ref{Pxt1nt2})
for the joint probability of the outcomes, as $P(n,t_2)\cdot
P(x,t_1|n,t_2)$ to emphasize the conditioning on the later photon
counting rather than the earlier homodyne detection event. Indeed,
such retrodictive conditioning can be applied to any type of
measurements, \textit{cf}. the past quantum state formalism \cite{PQS}, applied in
\cite{MURCH, DOTSENKO}. We shall briefly recall this formalism before
specializing to the calculation and interpretation of
amplitude--intensity correlations.

The density matrix evolves according to a completely positive map, and
hence, for any operator $\mu$, we can formally write the action of the
propagator of the master equation as a Kraus map \cite{WISEMAN, Nielsen},
$\mathrm{e}^{\mathcal{L}(t_2-t_1)}[\mu] = \sum_\alpha K_\alpha \mu
K^\dagger_\alpha$, where $\sum_\alpha K^\dagger_\alpha K_\alpha =
I$. The operators $K_\alpha$ represent the unobserved time evolution
between $t_1$ and $t_2$, and they depend of course on the explicit
form of the master equation and the propagation time, but for our
purpose we shall not need their explicit form. The cyclic properties
of the trace allows us to transform equation (\ref{jointdistribution})
into
\begin{equation} \label{twotimejointE}
P(m_1,t_1;m_2,t_2) = \textrm{Tr}(\hat{\Omega}_{m_1} \rho^{\mathrm{st}} \hat{\Omega}^\dagger_{m_1} \sum_\alpha K^\dagger_\alpha \hat{\Omega}^\dagger _{m_2}\hat{\Omega}_{m_2}  K_\alpha).
\end{equation}
It is now natural to define the operator product $E(t_1) \equiv
\sum_\alpha K^\dagger_\alpha \hat{\Omega}^\dagger
_{m_2}\hat{\Omega}_{m_2} K_\alpha$. The order of the
operators $K^\dagger_\alpha$ and $K_\alpha$ has been swapped compared
to their action on the time dependent density matrix, and the continuous equation of evolution for $E(t)$ is similarly obtained by swapping the Lindblad operator terms (\ref{lindblad}) in the master equation (\ref{MasterEq}),
\begin{equation}  \label{adjoint}
\frac{\mathrm{d}E(t)}{\mathrm{d}(-t)} = \frac{\mathrm{i}}{\hbar}[H,E] + \sum_n \hat{C}^\dagger_n E \hat{C}_n -  \frac{1}{2}(\hat{C}_n^\dagger \hat{C}_n E +
E \hat{C}_n^\dagger \hat{C}_n).
\end{equation}
We solve the equation for $E(t)$ backwards in time from $t_2$ to $t_1$ with the boundary condition, $E(t_2)=\hat{\Omega}^\dagger _{m_2}\hat{\Omega}_{m_2}$.

The correlations in the joint probability distribution imply that knowledge of the later measurement outcomes alters the conditional probability for the earlier ones. We can formally express this by $P(m_1,t_1;m_2,t_2)=P(m_2,t_2)\cdot P(m_1,t_1|m_2,t_2)$. We no´w determine the first factor $P(m_2,t_2)$ by summation over the index $m_1$ in (\ref{twotimejointE}) and the conditional probability then acquires the form
\begin{equation} \label{pqs}
P(m_1,t_1|m_2,t_2)=\frac{\mathrm{Tr}(\hat{\Omega}_{m_1} \rho(t_1) \hat{\Omega}^\dagger_{m_1} E(t_1))}{\sum_{m} \mathrm{Tr}(\hat{\Omega}_{m} \rho(t_1) \hat{\Omega}^\dagger_{m} E(t_1))},
\end{equation}
where the matrix $E(t_1)$ and the density matrix $\rho(t_1)$ play symmetric roles, representing how the outcome probabilities are correlated with the prior and the posterior dynamics and measurements of the system, respectively \cite{PQS}. We are interested in steady state correlation functions, and
since we eventually want to account for only the two-time correlation
with a single later photon detection event,
$\rho(t_1)=\rho^{\mathrm{st}}$. When we solve equation
(\ref{adjoint}), we obtain predictions in full agreement with the quantum regression theorem
results.

As $E(t)$ solves an equation which is similar to the master equation
for $\rho(t)$, we can exploit our intuition for density matrix
evolution and infer that for the two-level atom, $E(t)$ performs an
evolution backward in time, starting from the excited state, which is
similar to damped Rabi oscillations of the conventional density
matrix. This, indeed, gives a qualitative explanation of the red
dashed curve in figure \ref{2levelfig}. The evolution of $E(t)$
involves population loss out of the excited state by the Lindblad
damping term $-\frac{1}{2}(\hat{C}^\dagger \hat{C} E + E
\hat{C}^\dagger \hat{C})$ with $\hat{C}=\sqrt{\gamma}\hat{\sigma}$, but also population feeding into the excited state by
$\hat{C}^\dagger E \hat{C}$. Unlike the master equation
(\ref{MasterEq}), equation (\ref{adjoint}) does not preserve the
trace, and unlike the density matrix converging towards its steady
state value, $\rho^{\mathrm{st}}$, $E(t)$ converges towards the
identity matrix for long (negative) propagation times. It is thus not evident
that the correlation function in figure \ref{2levelfig} should be
completely symmetric in the time argument, but it follows by a closer inspection (and solution) of the equations of evolution.

%\begin{figure}
%\includegraphics[scale=0.3]{drawing}
%\caption{Measurement setups under consideration: (a) photocurrents cross-corelation corresponding to the intensity-intensity correlation function (b) homodyne detection}\label{CorrSchemes}
%\end{figure}

\section{Amplitude--intensity correlations from three-level atoms}
\label{sec:Three-level-atoms}
We shall in this section study the fluorescence from optical transitions in three-level atoms. These systems have alrady been analyzed \cite{CarmichaelREF2, CarmichaelREF3, CarmichaelREF1, CarmichaelREF4, Marquina-Cruz}, by the quantum regression theorem, and they show a number of features, that we can explain with our theory.
\subsection{Ladder system}

\begin{figure}[tbp]
\centering%
\includegraphics[bb=164 549 440
736,width=0.5\columnwidth,keepaspectratio]{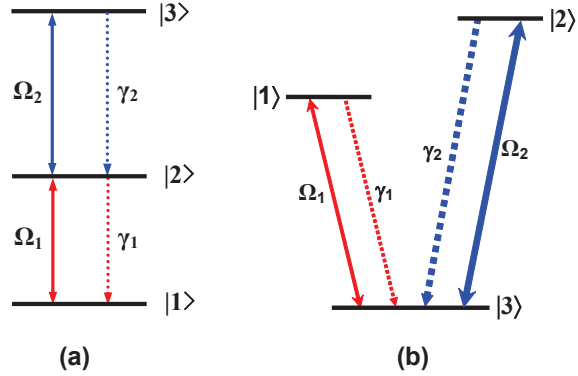}
\caption{A three-level atom driven by two resonant laser fields. In the ladder configuration (a), the fields are of comparable strength while the upper state $|3\rangle$ has longer lifetime than the intermediate state $|2\rangle$. In the $V$-configuration (b), the driving field and the decay rate on the $|2\rangle \leftrightarrow |3\rangle$ transition are much stronger than those on the $|1\rangle \leftrightarrow |3\rangle$ transition.}
\label{ladder-V}
\end{figure}

Let us consider a three-level ladder atom with energy levels as drawn in figure \ref{ladder-V}(a). Its Hamiltonian can be written as
\begin{equation} \label{ladderHamiltonian}
H =-\frac{\hbar }{2}(\Omega _{1}\hat{\sigma} _{21}+\Omega _{2}\hat{\sigma} _{32}+\mathrm{H.c.}),
\end{equation}
where $\hat{\sigma} _{ij}=|i\rangle \langle j|$, ($ i,j=1-3$), $\Omega
_{1,2}$ are the (real) laser Rabi frequencies. Quantum jump operators
of \eref{lindblad} are $\hat{C}_{1}=\sqrt{\gamma _{1}}\hat{\sigma}
_{12}$ and $\hat{C}_{2}=\sqrt{%
  \gamma _{2}}\hat{\sigma} _{23}$ with decay rates $\gamma _{1,2}$,
and $\hat{C}_{3}=\sqrt{\gamma _{\mathrm{ph}}}(\hat{\sigma}
_{33}-\hat{\sigma} _{22}-\hat{\sigma} _{11})$ with a dephasing rate
$\gamma _{\mathrm{ph}}$ of the upper state with respect to the two
lower states.

\begin{figure}[tbp]
\centering%
\includegraphics[bb=146 427 439
694,width=0.5\columnwidth,keepaspectratio]{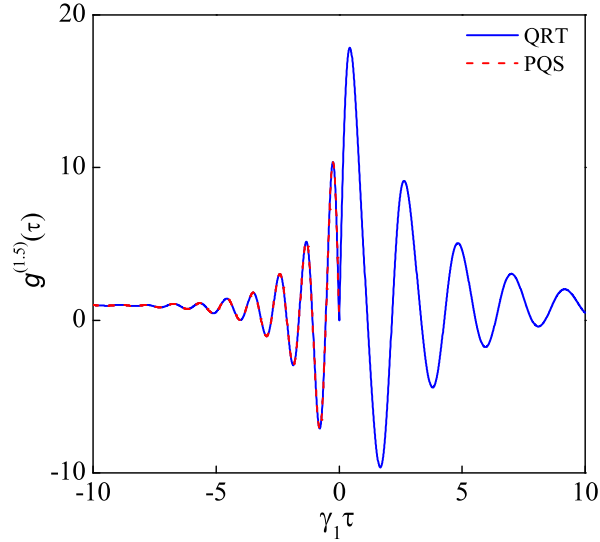}
\caption{(Color online) Amplitude--intensity correlation on the lower
transition of a single three-level ladder atom by invoking the quantum regression theorem (blue solid line) and past quantum state (red dashed line). The
parameters are: $\Omega _{1}=3\protect\gamma _{1}$, $\Omega _{2}=5\protect%
\gamma _{1}$, $\protect\gamma _{2}=0.2\protect\gamma _{1}$%
, $\protect\gamma _{\mathrm{ph}}=5\times 10^{-5}\protect\gamma _{1}$, and $%
\protect\varphi =\protect\pi /2$.}
\label{ladder-fig}
\end{figure}

We have calculated the correlation function $g^{(1.5)}$ between the
intensity and amplitude signals in the frequency range of both the
lower and the upper transition in the system. The quantum regression theorem, the POVM
formulation, and the past quantum state formalism naturally all give the same
quantitative predictions for this correlation function, shown in
figure \ref{ladder-fig}.

Unlike the two-level atom, the three-level atomic system shows a
striking temporal asymmetry of the correlation function
\cite{CarmichaelREF3}. In particular, for the lower transition the
frequency of the correlation function oscillations for the negative
time delay is twice the one for positive time delay (see figure
\ref{ladder-fig}). The past quantum state reasoning explains why this happens. For
$\tau > 0$, the photon counting event makes the atom density matrix
evolve forwards in time from the ground state $|1\rangle$ at time
$t$. For $\tau < 0$, the field amplitude can be calculated from the
matrix $E$, evolving backwards in time from the excited state of the
monitored transition, i.e., the intermediate state $|2\rangle$ in the ladder
configuration. In equations \eref{lindblad} and \eref{adjoint} the
evolution due to the Hamiltonian is the same (modulo a sign), and if
we disregard the dissipative terms, the evolution is readily analyzed
in terms of the Hamiltonian eigenstates,

\begin{eqnarray}
|+\rangle &=&\frac{1}{\sqrt{2}}(\cos \theta |1\rangle -|2\rangle +\sin
\theta |3\rangle ), \\
|0\rangle &=&-\sin \theta |1\rangle +\cos \theta |3\rangle ,
\label{dr1} \\
|-\rangle &=&\frac{1}{\sqrt{2}}(\cos \theta |1\rangle +|2\rangle +\sin
\theta |3\rangle ),
\end{eqnarray}
with corresponding eigenenergies
\begin{equation}
\mathcal{E}_{+}=\frac{\hbar }{2}\Omega _{\mathrm{R}},\quad \mathcal{E}_{0}=0,\quad \mathcal{E}_{-}=-\frac{%
\hbar }{2}\Omega _{\mathrm{R}},
\end{equation}
where  $\Omega _{\mathrm{%
R}}=\sqrt{\Omega _{1}^{2}+\Omega _{2}^{2}}$ and $\tan \theta =\frac{\Omega _{2}}{\Omega _{1}}$.

After a photon counting event, the conditioned evolution starts with
the atom in state $|1\rangle$, which is a superposition of all three
dressed eigenstates. It subsequently evolves with relative phase
factors that oscillate with the spectral separation
$\Omega_\mathrm{R}/2$, which then yields the oscillation frequency of
the density matrix elements. The backward evolution of $E(t)$, on the
other hand, starts from the state $|2\rangle$, which can be expanded as
a superposition of only the $|+\rangle$ and $|-\rangle$ eigenstates which
are separated by $ \Omega_\mathrm{R}$. The matrix elements of $E(t)$ thus
evolve with a relative frequency of $\Omega_\mathrm{R}$, which
therefore governs the twice faster oscillations of the two-time
correlation function for negative $\tau$.

\subsection{V-system}

The Hamiltonian of the $V$-configuration depicted in figure \ref{ladder-V}(b) reads
\begin{equation}
H =-\frac{\hbar }{2}(\Omega _{1}\hat{\sigma} _{13}+\Omega _{2}\hat{\sigma} _{23}+\mathrm{H.c.}),  \label{VeeHamiltonian}
\end{equation}
and the atom is subject to two quantum jump operators $\hat{C}_{1}=\sqrt{\gamma _{1}}\hat{\sigma} _{31}$ and $\hat{C}_{2}=\sqrt{%
\gamma _{2}}\hat{\sigma} _{32}$.

\begin{figure}[tbp]
\centering%
\includegraphics[bb=146 427 439
694,width=0.5\columnwidth,keepaspectratio]{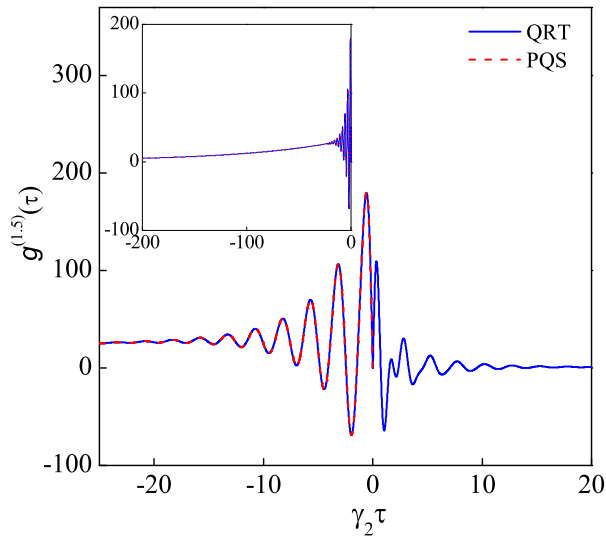}
\caption{Amplitude--intensity correlation on the weak probe
transition of a single three-level $V$ atom by invoking the quantum regression theorem (blue solid line) and
past quantum state (red dashed line). The inset indicates a long (negative) time trace approaching unity on the \textquotedblleft Zeno time scale\textquotedblright. The parameters are: $\Omega _{1}=0.1%
\protect\gamma _{2}$, $\Omega _{2}=5\protect\gamma _{2}$, $\protect\gamma _{1}=0.01\protect\gamma _{2}$, and $\protect\varphi =%
\protect\pi /2$. }
\label{Vee-fig}
\end{figure}

As shown in figure \ref{Vee-fig}, the amplitude--intensity correlation
function on the weak transition $ |1\rangle \leftrightarrow |3\rangle
$ in a $V$-system is asymmetric. For $\tau > 0$, we observe damped
dipole moment oscillations, with an additional modulation that we
ascribe to the rapid dynamics on the stronger atomic transition. For
$\tau < 0$, the oscillations are more regular, however they are damped
towards a large almost constant value, and only on a much longer time
scale (see insert), the correlation function converges to unity. We
can account for both of these observations by the backward evolution
of the matrix $E(t)$: First, we note that $E(t)$ has the excited state
$|1\rangle \langle 1|$ as its final state. This state is resonantly
driven towards the state $|3\rangle$, which is, however, strongly
perturbed by the laser excitation of the $|2\rangle \leftrightarrow |3\rangle$
transition with Rabi frequency $\Omega_2$. This driving causes an AC
Stark splitting of the state $|3\rangle$ by $\Omega_2$ and detunes the weak transition by $\Omega_2/2$,
which emerges as a generalized Rabi frequency
of the weak and oscillatory amplitude transfer. The incoherent rate
processes between states $|2\rangle$ and $|3\rangle$ damp these
off-resonant oscillations on the time scale of a few
$(\gamma_2)^{-1}$. The matrix $E(t)$ has not yet reached its steady
state value, and despite the coherent coupling strength $\Omega_1$, it
is only for times longer than the time scale $(\gamma_1)^{-1}$, that
the correlation function regresses to the uncorrelated product. Within
the picture of an evolving matrix $E(t)$, we find this
suppression of coherent dynamics similar to the quantum Zeno dynamics
\cite{Facchi}, observed for evolving density matrices
\cite{Shao}.

% for the explanation and we need to inspect the eigenvalues and eigenstates of the whole Liouvillian. Unlike previous cases we cannot make use of the suggestive dressed-state picture so we provide at least numerical study of the situation.
%\begin{figure}[tbp]
%\centering%
%%\includegraphics[scale=0.6]{Figure-7.pdf}
%\caption{(Color online) The imaginary parts versus the real parts of the
%eigenvalues associated with the evolution matrices of $\protect\rho $ matrix
%(blue circle) and $E$ matrix (red diamond). The parameters are: $\Omega
%_{1}=0.1\protect\gamma _{2}$, $\Omega _{2}=5\protect\gamma _{1}$, $\Delta
%_{1,2}=0$, and $\protect\gamma _{1}=0.01\protect\gamma _{2}$.}
%\label{eigenvalues}
%\end{figure}
% The eigenvalues of both Liouvillian for forward evolution and its adjoint counterpart are plotted in the Fig. \ref{eigenvalues}. First,let us note that the eigenvalues of the Liouvillian and adjoint Liouvillian are the same and remind that the eigenstate of the eigenvalue zero for Liouvilian is steady state, for adjoint Liouvillian it is the identity. Now, we see one additional eigenvalue that is very close to zero (-0.01â€¦ in the example), and we have found that the corresponding eigenvector of adjoint Liouvillian is almost a projector on the state $|2\rangle$. Hence, the time evolution effected by the adjoint master equation originated from state $|2\rangle$, will only contain the slow decay associated to the close-to-zero eigenvalue, since this state projects almost completely on the corresponding eigenvector.

\section{Conclusion}
\label{sec:conclusion}
Temporal correlation functions witness the non-classical character of
light emitted from quantum sources, and they have played an immense
role in the analysis and in the application of quantum states of
light. In this article we have revisited the calculation of such
correlation functions in terms of generalized measurements, and we
have discussed how the correlations can be interpreted in terms of
measurement back action and transient evolution of the emitter system
quantum state during continuous probing.

Transients after discrete photon counting events have been well
understood as the conventional state dynamics from the non-steady
state prepared by the measurement back action of the counting
event. To account for transients before such discrete measurement
events, we employed the theory of past quantum states which allows
calculation of measurement outcome probabilities, conditioned on the
knowledge of both earlier and later measurements on the system. The
practical calculations associated with this method are equivalent to
the ones of the conventional master equation and the quantum
regression theorem, but they deal explicitly with the transient
evolution from a certain initial (final) state, according to an
equation with enough similarities with the conventional master
equation to offer insight into the dynamics.

We applied the method to amplitude--intensity correlations in the
fluorescence from laser driven two- and three-level atoms. We recall,
that the pertaining correlation functions have already been calculated
in previous works, but without the interpretations offered by the
present analysis.

\bigskip\noindent
\textbf{Acknowledgments}

This work was supported by the Villum Foundation and the
IARPA MQCO program. EG acknowledges discussions with C. Navarrete-Benlloch and A. Gonzalez-Tudela.

\end{document}